# Improving the understandability of the next edition of the International System of Units (SI) by focusing on its conceptual structure

Luca Mari[*#], Peter Blattner[**], Franco Pavese[***]
* Università Cattaneo-LIUC, 21053 Castellanza (VA), Italy
** Federal Institute of Metrology-METAS, 3003 Bern-Wabern, Switzerland
*** Independent scientist[§]

## Abstract

The International System of Units (SI) is fundamental for the social, and not only the scientific, role of metrology, and as such its understandability is a crucial issue. The SI is officially presented in a document commonly called the "SI Brochure", published by the International Bureau of Weights and Measures (BIPM). According to the current draft of the new SI Brochure, the next edition of the SI will be significantly more complex in its conceptual structure than the previous ones. Identifying a strategy for effectively communicating its main contents is then a worthwhile endeavor, in order to increase the acceptance and thus the sustainability of the SI itself. Our proposal is to focus on the structure of the definitions, while omitting all physical contents, which are inaccessible to most potential readers: this is instrumental to the awareness campaigns recommended by the General Conference on Weights and Measures (CGPM) to make the next edition of the SI understandable by a diverse readership without compromising scientific rigor. By unpacking the structural contents, the structure of definitions that we finally propose has the merit not only of being understandable, but also of highlighting the fundamental nature of metrology: a complex body of knowledge intertwining science and technology, society, and language even in its fundamental definitions.

## 1. Introduction

Given the role that measurement plays in many aspects of our life for guaranteeing the public trust of values attributed to quantities [BIPM 2007], the widespread understanding not only of the basic structure of the metrological system as such but also of its grounding element – a socially agreed system of units [VIM, def. 1.13] – is a crucial issue. The key point is that even the best measuring system cannot provide reliable results if it was not properly calibrated, because it is the calibration against appropriate measurement standards that guarantees the metrological traceability of measurement results. The International System of Units (SI) [VIM, def. 1.16] is the system of units developed under the supervision of the General Conference on Weights and Measures (CGPM) and as such is almost universally adopted.

The SI is officially illustrated in a document titled "The International System of Units (SI)" and commonly referred to as the "SI Brochure", published by the International Bureau of Weights and Measures (BIPM), which operates under the exclusive supervision of the International Committee for Weights and Measures (CIPM), which itself comes under the authority of the CGPM.

Since several years a process of radical revision of the structure itself of the SI is ongoing, as witnessed in the drafts of the next, 9[th] edition of SI Brochure[1]. Such revised SI will be based on a

---

complex conceptual structure, in which quantity units[2] are indirectly defined by reference to constant quantities and by setting their values, thus reversing the conceptual sequence adopted so far, in which units have been defined first and from them constants have become measurable. While several papers already discussed this structure (e.g., [Milton et al 2007], [Mohr 2008], [Cabiati, Bich 2009], [Mills et al. 2011], [Milton et al 2014], [Newell 2014]), it is acknowledged – and in fact it has been authoritatively recommended[3] – that increasing the awareness on the next edition of the SI, thus supposedly also in terms of its understandability, is still an important task.

In this respect, the prompting issue here is that content-related details are just inaccessible without a solid scientific background, particularly in quantum physics: this is a knowledge available to a much more restricted community than the one interested in the SI and its use, that in industrialized countries practically coincides with the whole society. That pupils of primary schools and the canonical man-on-the-street can and will be exposed only to a very simplified presentation of the SI is to be taken for granted: outside technical environments people are seldom expected to learn through definitions, and their knowledge is usually acquired by example, thus with a bottom-up attitudeI. In the case of the SI, this might correspond to emphasizing on simple *realizations* of the units rather than their *definitions* (the specific ways of realization of the units – the so called *mise en pratique* – will not be discussed in this paper: see on this matter the draft of the SI Brochure [SI Brochure] and the related draft examples [BIPM 2016]).

On the other hand, assuming that everyone without a degree in physics should approach the SI as a black box would produce critical consequences, first of all to turn metrology away from society, given instead the importance that science and technology are more and more explicitly presented as fundamental components of our society and therefore widely understood at least in their basic principles. Challenging is then the endeavor of effectively disseminating the conceptual framework that grounds the next edition of the SI to high school and university students, technicians in companies and calibration laboratories, etc. While most scientific information related to the physical content might remain out of reach, they could appreciate the elegant conceptual structure of the framework, and through it not only grasp the fundamentals of the SI but also better understand such critical concepts as physical constant and quantity unit, and eventually be able to justify metrological traceability [VIM, def. 2.41] as the actual basis of the public trust attributed to measurement results. This requires that the main contents of the next edition of the SI are presented in a structural perspective able to provide simple and sound information, and that definitions are explained in a recognizably consistent conceptual framework and corresponding linguistic format. The presentation of such a conceptual structure is the main target of the present paper.

---

1  All references to the next edition of the SI Brochure are based on its current draft [SI Brochure], published in the web site of the BIPM on November 2016.

2  While the historically established term is "measurement unit" (as defined in the International Vocabulary of Metrology [VIM]), there are at least two reasons to consider that "quantity unit" is a better term. First, units apply not only to measurements, but more generally to all situations in which values of quantities are involved; second, the expression "quantity unit" more correctly can be specified as, e.g., "length unit", or "unit of length", where length is a (kind of) quantity, not a measurement. Almost everywhere the SI Brochure simply uses "unit", and we will do the same here.

3  "The General Conference on Weights and Measures (CGPM), at its 24th meeting, [...] invites [...] the CIPM, the Consultative Committees, the BIPM, the OIML and National Metrology Institutes significantly to increase their efforts to initiate awareness campaigns aimed at alerting user communities and the general public to the intention to redefine various units of the SI and to encourage consideration of the practical, technical, and legislative implications of such redefinitions, so that comments and contributions can be solicited from the wider scientific and user communities." [CGPM 2011]. "The General Conference on Weights and Measures (CGPM), at its 25th meeting, [...] noting that further work by the Consultative Committee for Units (CCU), the CIPM, the BIPM, the NMIs and the CCs should focus on (i) awareness campaigns to alert user communities as well as the general public to the proposed revision of the SI, (ii) the preparation of the 9th edition of the SI Brochure that presents the revised SI in a way that can be understood by a diverse readership without compromising scientific rigour, [...] encourages ..." [CGPM 2014].

As a starting point let us consider the definition that the current draft in the SI Brochure [SI Brochure, sec. 2.2.1] gives of the unit of time[4], structurally the simplest one:

> "The second, symbol s, is the SI unit of time. It is defined by taking the fixed numerical value of the caesium frequency $\Delta v_{Cs}$, the unperturbed ground-state hyperfine transition frequency of the caesium 133 atom, to be 9 192 631 770 when expressed in the unit Hz, which is equal to s$^{-1}$."

The reader finds here two features that s/he learned a "good" definition should *not* have: (i) the definition includes the defined concept (this is in fact the definition of what the second is – so that, at least in principle, before such a definition the concept 'second' would just be lacking – but the second is defined here in terms of the second); (ii) the definition defines a concept in terms of previously undefined concepts, without assuming them as primitives (the second is defined in terms of hertz).

The point here is not to argue once again on the correctness of this kind of definitions, not to propose any change of them: even though the current draft of the SI Brochure might be refined before its final publication, we will take it for granted here. Rather, the issue is about the strategy to make the new definitions understandable to the widest community as possible. This paper proposes an explicit presentation by unpacking the structural components of the definitions. What follows is then a proposal instrumental to the awareness campaigns recommended by the General Conference on Weights and Measures to make the next edition of the SI understandable "by a diverse readership without compromising scientific rigour" [CGPM 2014] (see a previous footnote). Accordingly, the discussion is not devoted to any specific target group (teachers, university students, technicians, etc), and can be intended as a conceptual template that can be variously customized and enriched with notes and examples in function of the given target.

The paper is structured as follows. Section 2 starts from the simplest possible meaningful subset of the SI and unpacks it, thus revealing its fundamental structure, defining what could be called a Fundamental System of Units. From such a system, Sections 3 and 4 reconstruct the SI by showing that the added complexity in it is required to satisfy two socially critical *principles of continuity*. Section 5 shortly discusses the trade-off between complexity and continuity as a strategic decision in the presentation of the next edition of the SI. The Appendix provides the conceptual and lexical background for introducing our subject.

## 2. The fundamental structure of a constant-based system of units

At its core the next edition of the SI can be seen as making an inversion of conceptual priority: while until now some quantities of objects or physical states have been conventionally adopted as units and from them the values of physical constants have been measured in reference to such predefined units, in the new scenario physical constants and their values come first and from them units are deduced. There are three sources of complexity in this strategy:

1. the physical nature of the constant quantities involved in the definitions;
2. the fact that such constants are linked to multiple kinds of quantities and therefore that they interdefine multiple units;
3. the inversion of conceptual priority as such.

While the first source cannot be amended, the presentation may be made simpler by taking at first only a subset of constants into account, thus reducing the complexity due to the second source and making it possible to focus on the third source. In this perspective the simplest option is to elaborate only on the caesium frequency constant $\Delta v_{Cs}$, and therefore from it on the second as the unit of duration. According to the draft SI Brochure [sec. 2.1]:

---

4  The reference should actually be to duration, indeed, not time, which is not a quantity. Exactly in the same sense the metre is, correctly, the unit of length, not space.

> The International System of Units, the SI, is the system of units in which the unperturbed ground state hyperfine transition frequency of the caesium 133 atom $\Delta v_{Cs}$ is 9 192 631 770 Hz [etc]

which can be simplified, by omitting here immaterial phraseology, as:

D0. The SI is the system of units in which the frequency $\Delta v_{Cs}$ is taken as 9 192 631 770 Hz

The first difficulty in this definition is that it includes a reference to a unit, hertz, that is still undefined. Hence translating it in the symbolic version:

$\Delta v_{Cs} := 9\,192\,631\,770$ Hz                                         // wrong

is not correct[5], due to the fact that the right-hand side term in a definition (the *definiens*) must include only predefined or primitive concepts. Moreover, $\Delta v_{Cs}$ is the quantity of a kind of object, assumed as constant according to the best available physical theories, that as such it does not require any definition. While still not correct, a simple solution to the first issue would be to rewrite definition D0 as follows:

D1. The SI is the system of units in which the frequency $\Delta v_{Cs}$ is taken as 9 192 631 770 units of frequency

i.e.:

$\Delta v_{Cs} := 9\,192\,631\,770$ units of frequency                        // almost correct

where the concept 'unit of frequency' is a placeholder for a still undefined concept. This highlights the inverse structure of the definition, which may be made more explicit by further rewriting it:

D2. The SI is the system of units in which the unit of frequency is 1/9 192 631 770 the frequency $\Delta v_{Cs}$

There is no harm at this point in introducing a specific term or symbol for the unit of frequency appearing in the left-hand side term of the definition (the *definiendum*):

D3. The SI is the system of units in which the unit of frequency, the hertz, symbol Hz, is 1/9 192 631 770 the frequency $\Delta v_{Cs}$

so that:

Hz := 1/9 192 631 770 $\Delta v_{Cs}$                                         // correct

is perfectly admissible, being an instance of the relation (4) in the Appendix.

As the final step of this deconstruction, we can acknowledge that a numerical value such as 9 192 631 770 does not have any structural role in the system, being introduced to guarantee that the quantity identified as unit does not change when its definition changes. Hence, in principle we can decouple the structure from its historical constraints, and propose the explicit introduction of a system of units – let us call it the "Fundamental System of Units"[6] – in which the defined units are related to the defining constants by the factor 1, and therefore such that:

D4. The Fundamental System of Units is the system of units in which the frequency $\Delta v_{Cs}$ is taken as the unit of frequency i.e., the unit of frequency is the frequency $\Delta v_{Cs}$.

We might call the unit of frequency in the Fundamental System of Units "fundamental-hertz", $Hz_f$, and then:

$Hz_f := \Delta v_{Cs}$

a definition that has the merit of being structurally analogous to the traditional definitions of quantity units (the hertz is the frequency of the object such that…; the kilogram is the mass of the object such that..., etc), and therefore easily understandable in its structure.

---

5  See the Appendix about the difference between empirical equality and equality by definition, which is critically important here.

6  An alternative name could be "natural system of units" (see, e.g., https://en.wikipedia.org/wiki/Natural_units), but we prefer avoiding it because (i) multiple versions of it have been proposed and (ii) the adjective "natural" conveys the unclear, if not misleading, message that all other systems of units, the SI included, are in some sense non-natural.

The explicit structure of a definition in the Fundamental System of Units is then [Mari, Pavese 2016]:

D5.  The Fundamental System of Units is the system of units in which:

   (i) the frequency $\Delta\nu_{Cs}$ is taken as the unit of frequency, the fundamental-hertz, symbol $Hz_f$

   (ii) then the frequency $\Delta\nu_{Cs}$ is 1 $Hz_f$

The simplicity of the Fundamental System of Units is maintained even when the units of other kinds of quantities related to the defining constants are introduced, for example, together with frequency, speed and action:

D6.  The Fundamental System of Units is the system of units in which:

   (i) the frequency $\Delta\nu_{Cs}$ is taken as the unit of frequency [frequency]$_f$, the speed $c$ is the unit of speed [speed]$_f$, the action $h$ is the unit of action [action]$_f$

   (ii) then the frequency $\Delta\nu_{Cs}$ is 1 [frequency]$_f$, the speed $c$ is 1 [speed]$_f$, the action $h$ is 1 [action]$_f$

where, in the lack of accepted terms for the units of speed and action, the units in the Fundamental System of Units have been denoted as [*quantity*]$_f$.

This simplicity is obtained at the price of accepting that:

– the units in the Fundamental System of Units and the SI units are different quantities (for example, the fundamental-hertz and the hertz are different quantities by a factor 9 192 631 770);

– the kinds of quantities for which the units are defined are not the base quantities in the SI (for example, the Fundamental System of Units defines a unit for frequency instead of duration)[7].

There are two socially critical *principles of continuity*[8] at stake:

PRINCIPLE 1: the quantities identified as units should remain the same even when their definition changes;

PRINCIPLE 2: the units should be defined of the kinds of quantities that have been assumed as base quantities so far,

and both are violated here[9]. The consequence is that the Fundamental System of Units might be even considered excellent by theoretical physicists but would be impractical in most social situations. This shows that the complexity added in the SI starting from the Fundamental System of Units is a price to be paid for the social acceptance of the SI itself.

## 3. The definition of a SI without base units

A possible first step from the Fundamental System of Units toward the SI takes PRINCIPLE 1 into account: this preserves the set of units of the present SI but still does not define base units, as instead PRINCIPLE 2 would require.

By extending definition D1 to the case of multiple kinds of quantities (here in reference to a system for units of duration, length, and mass for the sake of simplicity), it becomes:

D1'.  The SI is the system of units in which the frequency $\Delta\nu_{Cs}$ is 9 192 631 770 units of frequency, the speed $c$ is 299 792 458 units of speed, and the action $h$ is $6.626\,069\,3 \times 10^{-34}$ units of action

so that the rewriting of definition D2 is immediate:

D2'.  The SI is the system of units in which the unit of frequency is 1/9 192 631 770 the frequency $\Delta\nu_{Cs}$, the unit of speed is 1/299 792 458 the speed $c$, and the unit of action is $(1/6.626\,069\,3) \times 10^{34}$ the action $h$

---

7  Of course, we are using here the terminology of the International Vocabulary of Metrology [VIM], such that an entity like $\Delta\nu_{Cs}$ is a quantity, of the kind frequency.

8  The continuity in time of the units through the changes of definitions is also called "constancy" [Johansson 2015], since it guarantees the constancy in time of the numerical values of the relevant quantities.

9  "Preserving continuity, as far as possible, has always been an essential feature of any changes to the International System of Units." [SI Brochure, sec. 2.1].

The actual content of definition D2' is indeed [Newell 2014]:

$$[\text{frequency}] := k_{\Delta\nu_{Cs}}^{-1} \Delta\nu_{Cs}$$
$$[\text{speed}] := k_c^{-1} c$$
$$[\text{action}] := k_h^{-1} h$$

where the fact that $k_{\Delta\nu_{Cs}}$, $k_c$, and $k_h$ appear in their inverse form highlights the indirect structure of these definitions: first the constants, then the units, which is the very basic intention of the revision of the SI.

Definition D2' is then unpacked as:

D6'. The SI is the system of units in which:

(i) the unit of frequency is [frequency], the unit of speed is [speed], the unit of action is [action]

(ii) the frequency $\Delta\nu_{Cs}$ is $k_{\Delta\nu_{Cs}}$ [frequency]$_f$, the speed $c$ is $k_c$ [speed]$_f$, the action $h$ is $k_h$ [action]$_f$

(iii) then [frequency] is the frequency $k_{\Delta\nu_{Cs}}^{-1} \Delta\nu_{Cs}$, [speed] is the speed $k_c^{-1} c$, [action] is the action $k_h^{-1} h$

(iv) where the numerical value $k_{\Delta\nu_{Cs}}$ is 9 192 631 770, the numerical value $k_c$ is 299 792 458, and the numerical value $k_h$ is $6.626\,069\,3 \times 10^{-34}$ where the basic difference with respect to definition D6 is the presence of numerical values different from 1.

This definition is still remarkably simple in its structure[10], thanks to the fact the defined units are quantities of the same kind as the defining constants: by fulfilling PRINCIPLE 1 the clauses (i-iii) add negligible complexity to the structure, only related, in the clause (iv), to the presence of the conversion factors $k \neq 1$. Based on this strategy a system of units can then be built up without introducing the distinction between base and derived units.

## 4. The structure of the definitions in the revised SI

The next edition of the SI can be intended as the outcome of the changes introduced by the Fundamental System of Units, as synthesized in definition D6, in order to keep both mentioned PRINCIPLEs of continuity into account. Taking also PRINCIPLE 2 into account requires to maintain duration, length, and mass as base quantities, and therefore to revise definition D6' in order to define the units of such kinds of quantities instead of those of frequency, speed, and action. To this goal physics becomes unavoidable, via quantity equations, i.e., "mathematical relations between [kinds of] quantities in a given system of quantities, independent of measurement [i.e., quantity] units"[11] [VIM, def. 1.22], e.g.:

$$\text{duration} = \text{frequency}^{-1}$$

On this basis, the definition becomes finally:

D6''. The SI is the system of units in which:

(i) the base unit of duration is the second, symbol s, the base unit of length is the metre, symbol m, and the base unit of mass is the kilogram, symbol kg;

(ii) the frequency $\Delta\nu_{Cs}$ is $k_{\Delta\nu_{Cs}}$ s$^{-1}$, the speed $c$ is $k_c$ m s$^{-1}$, the action $h$ is $k_h$ kg m$^2$ s$^{-1}$;

(iii) then the second is $k_{\Delta\nu_{Cs}} \Delta\nu_{Cs}^{-1}$, the metre is $k_c^{-1} c$ s, and the kilogram is $k_h^{-1} h$ s m$^{-2}$;

(iv) the numerical value $k_{\Delta\nu_{Cs}}$ is 9 192 631 770, the numerical value $k_c$ is 299 792 458, and the numerical value $k_h$ is $6.626\,069\,3 \times 10^{-34}$ which fulfills both PRINCIPLE 1 and PRINCIPLE 2. The complexity of this definition is apparent in particular in the clause (iii), which in fact could be rewritten, by substitution, as:

$$s := k_{\Delta v_{Cs}} \Delta v_{Cs}^{-1}$$
$$m := k_{\Delta v} Cs \, k_c^{-1} \Delta v_{Cs}^{-1} c$$
$$kg := k_c^2 \, k_{\Delta v} Cs^{-1} \, k_h^{-1} \Delta v_{Cs} \, c^{-2} h$$

all of them being instances of the relation (4) in Appendix. Hence, *the base units are expressed as products of powers of the defining constants*.

---

[10] The assignment of the best values to $k_{\Delta\nu_{Cs}}$, $k_c$, and $k_h$ is the delicate task of CODATA, which, in this stage of transition from the current to the next version of the SI, is basically to guarantee the fulfillment of PRINCIPLE 1. The structure of the definition D6' is then parametric: the clauses (i-iii) specify the structure, and the clause (iv) sets the values of the parameters.

[11] The independence of quantity equations from quantity units is clearly of utmost importance here.

It is clear that the presence of the numerical constants $k$, aimed at fulfilling PRINCIPLE 1, is only a marginal source of complexity here: much more critical is the choice of maintaining the units of duration, length, and mass as base units and fixing their numerical values accordingly.

## 5. Discussion and conclusions

In this paper three different systems of units have been presented:

1. the Fundamental System of Units (D6), which can considered excellent for theoretical physics but would be very impractical in most social situations;
2. the SI without the traditional base units (D6');
3. the SI as proposed in the 9th edition of the SI Brochure (D6'').

The following table compares D6, D6', and D6'':

| definition | relative degree of complexity | relative degree of continuity |
| --- | --- | --- |
| D6: Fundamental System of Units | low | low: neither PRINCIPLE 1 nor PRINCIPLE 2 are fulfilled |
| D6': SI without traditional base units | medium | medium: only PRINCIPLE 1 is fulfilled |
| D6'': SI, as proposed in the 9th edition of the SI Brochure | high | high: both PRINCIPLE 1 and PRINCIPLE 2 are fulfilled |

While this is not the context to argue about the opportunity to pay the price of the increased complexity of definition D6'' for maintaining the traditional base quantities, a simple note can be proposed on this matter. As mentioned, the identification of base quantities is no more related to the SI units, and is instead grounded on some substantial conditions (in particular the mutual independence of the base quantities, in the sense that "a base quantity cannot be expressed as a product of powers of the other base quantities" [VIM, def. 1.4 n.2]) together with considerations of conceptual simplicity (according to the way physics is taught and learned, for example length is usually intended as "more fundamental" than speed). This seems to be more than enough to maintain the current International System of Quantities (ISQ), "system of quantities based on the seven base quantities: length, mass, time, electric current, thermodynamic temperature, amount of substance, and luminous intensity" [VIM, def. 1.6], on which quantity calculus / dimensional analysis has been traditionally based.

The definition of 'base unit', "measurement [i.e., quantity] unit that is adopted by convention for a base quantity" [VIM, def. 1.10] rightly shows the dependence of base units on base quantities, not on any particular structure of unit definition. Hence, the simpler definition D6' could be adopted to present the next edition of the SI in the awareness campaigns, together with a possible note showing how the traditional base units (the second, etc) are derived from the new ones.

In this perspective a definition such as:

D7. The SI is the system of units in which:

  (i) the unit of frequency is the hertz, symbol Hz

(ii) the frequency $\Delta v_{Cs}$ is $k_{\Delta v_{Cs}}$ Hz

(iii) then the hertz is $k_{\Delta v_{Cs}}^{-1} \Delta v_{Cs}$

(iv) for continuity reasons the numerical value $k_{\Delta v_{Cs}}$ is chosen to be 9 192 631 770

Note: the base unit of duration in the SI is the second, symbol s, defined as $Hz^{-1}$

is the simplest unpacked version of a definition based on the new structure. Definitions D7 and D6'' have an analogous structure[12], which is particularly interesting:
– the clause (i) is lexical, and introduces a term and a symbol for convenience;
– the clause (ii) conveys the core empirical knowledge on the quantity whose unit is under definition, and on this basis the clause (iii) defines the unit;
– the clause (iv) implements PRINCIPLE 1;
– the note shows how also PRINCIPLE 2 can be substantially maintained.
Such a structure has the merit not only of being effectively understandable, but also of highlighting the fundamental nature of metrology: a complex body of knowledge intertwining science and technology, society, and language even in its fundamental definitions.

## Appendix: Quantities of objects and values of quantities

At the basis of the understanding of the role of quantity units there is the complex relation between quantities of objects, such as the length of a given pen, and values of quantities, such as 0.123 m. On the one hand, such two types of entities are conceptually distinct: the pen has a physical property, its length, that is independent of the definition of any quantity unit and therefore of quantity values; the quantity value 0.123 m can be identified independently of the existence of any physical object having that length. On the other hand, an experimental activity could lead to assess that the length of the pen is 0.123 m (measurement uncertainty will not be considered here), thus producing the information that two conceptually distinct entities are in fact the same[13].
Acknowledging then that
(i) the difference between quantities of objects and values of quantities is in the type of knowledge that we can have of them, and that
(ii) quantities of objects can be modeled as variables taking values of quantities as their values,
the usual statistical notation can be adopted: the variable is denoted by an uppercase character, say $Q$, and its value by the corresponding lowercase character, $q$. An experimental activity could then lead to assess that:

$$Q = q \qquad (1)$$

thus meaning that (i) while we have different knowledge on a quantity of an object (e.g., the length of this pen), $Q$, and a value of a quantity (e.g., 0.123 m), $q$, and that (ii) we are claiming that they are the same. Precisely this twofoldness – different but equal – makes the relation non-conventional and informative: still neglecting the role of uncertainty, it is indeed either true or false.

---

[12] This analogy does not hide a delicate point in definition D7: since according to the SI Brochure [sec. 2.2.4, table 4] the hertz is a derived unit, one might conclude that the second is defined in terms of the hertz (s := $Hz^{-1}$) and then the hertz is defined in terms of the second (Hz := $s^{-1}$). In our view this is one more reason for preferring the structure of D6'' to present the conceptual bases of the next edition of the SI.

[13] The discovery that conceptually different entities are the same is an example of the knowledge advancement expected from measurement, and a canonical case of the complexity of the relation of equality (which is not identity in all possible respects). In a famous paper on this subject [Frege 1892], G. Frege proposed the example of relation between the morning star and the evening star: while conceptually distinct, they were finally discovered to be the same celestial body. Sameness is more controversial in the case of quantities of objects and values of quantities, given that one could assume that, e.g., the length of this pen is a concrete / particular entity and 0.123 m is an abstract / universal one. In this view, the usual expression "length(this pen) = 0.123 m" could be meant length(this pen) Î 0.123 m, where then values of quantities are intended as classes of quantities. This controversy is not important here, and is further discussed in [Mari, Giordani 2012].

In the case of kinds of quantities such as length[14], any value is built by assuming that a given quantity, taken as the unit, is additively replicable a given number of times. This can be expressed using the Maxwell's notation:

$$q := \{Q\}[Q] \qquad (2)$$

where $\{Q\}$ is a numerical quantity value and $[Q]$ is a quantity unit.

It is important to note that $:=$ denotes here an is-defined-as relation, and therefore a conventional, non-empirical, and non-symmetrical relation, which is neither true nor false.

By substituting (2) in (1) we obtain the well-known:

$$Q = \{Q\}[Q] \qquad (3)$$

that, in the same sense as (1), is non-conventional and informative, and either true or false.

In order to further emphasize the difference between (2) and (3), consider that (3) is an equality, on which the usual transformations are allowed, in particular:

$$Q/[Q] = \{Q\}$$

that is a specific instance of the Euclidean characterization of numbers as ratios of quantities. On the contrary, the fact that (2) is a definition prevents analogous manipulations, for example by transforming it into:

$$q/[Q] := \{Q\} \qquad \text{// wrong}$$

which is clearly wrong (just as another example, the relation is-defined-as is not symmetric: of course, from $x := y$ it does not follow that $y := x$). On the other hand, a relation such as:

$$[Q] := k\, Q \qquad (4)$$

where $k$ is a numerical constant, is perfectly admissible, meaning that the unit $[Q]$ is going to be defined as the $k$-multiple of the quantity of an object, $Q$, that is assumed to be known (it should obvious then that $k\,Q$ in (4) is not a value). Like (2), also (4) is conventional and non-empirical, and neither true nor false.

Our claim is that this is the basis for discussing how the definitions in the draft SI Brochure can be made effectively understandable to readers without any knowledge of quantum physics[15].

**Acknowledgments**

---

14  Possible terms for a kind of quantity such as length are "Euclidean quantity", "ratio quantity", "ratio scale quantity", "additive quantity", "unitary quantity", etc: while not necessarily conveying exactly the same meaning, the difference is not important here.

15  While this paper aims at proposing materials the complement, not modify, the SI Brochure, what has been proposed so far shows also that some details in the current draft of the SI Brochure could be improved, in particular by using distinct symbols for the relations is-defined-as and is-equal-to and distinct symbols for quantities of objects and values of quantities. Furthermore, the idea that "the unit is a particular example of the value of a quantity" (at the beginning of Section 2) seems to be just wrong: as the International Vocabulary of Metrology [VIM] clearly defines, units are quantities of objects, not values.